\documentclass[aps,pre,preprint,showpacs,superscriptaddress]{revtex4-1}
\usepackage{amsmath}
\usepackage{amssymb}
\usepackage{gensymb}
\usepackage{graphicx}
\usepackage{epstopdf}
\usepackage[english]{babel}
\usepackage{siunitx}

\usepackage{xcolor}
\usepackage{hyperref}
\definecolor{citecolor}{rgb}{0,0,0}
\hypersetup{colorlinks=true,
	linkcolor={citecolor},
	citecolor={citecolor},
	urlcolor={citecolor}}
\begin{document}

\title{Evidence of a first-order smectic -- hexatic transition and its proximity to tricritical point in smectic films}

\author{Ivan~A.~Zaluzhnyy}
\affiliation{Deutsches
	Elektronen-Synchrotron DESY, Notkestra{\ss}e 85, D-22607 Hamburg,
	Germany}
\affiliation{National Research Nuclear University MEPhI (Moscow Engineering Physics Institute), Kashirskoe shosse 31, 115409 Moscow, Russia}

\author{Ruslan~P.~Kurta}
\affiliation{European XFEL GmbH, Holzkoppel 4, D-22869 Schenefeld, Germany}

\author{Nastasia~Mukharamova}
\affiliation{Deutsches Elektronen-Synchrotron DESY, Notkestra{\ss}e 85, D-22607 Hamburg, Germany}

\author{Young~Yong~Kim}
\affiliation{Deutsches Elektronen-Synchrotron DESY, Notkestra{\ss}e 85, D-22607 Hamburg, Germany}

\author{Ruslan~M.~Khubbutdinov}
\affiliation{Deutsches Elektronen-Synchrotron DESY,
	Notkestra{\ss}e 85, D-22607 Hamburg, Germany}
\affiliation{National Research
	Nuclear University MEPhI (Moscow Engineering Physics Institute),
	Kashirskoe shosse 31, 115409 Moscow, Russia}
\affiliation{Institute of Solid State Physics, Russian Academy of
	Sciences, Academician Ossipyan str. 2, 142432 Chernogolovka, Russia}

\author{Dmitry~Dzhigaev}
\affiliation{Deutsches Elektronen-Synchrotron DESY, Notkestra{\ss}e 85, D-22607 Hamburg, Germany}

\author{Vladimir~V.~Lebedev}
\affiliation{Landau Institute for Theoretical Physics, Russian
	Academy of Sciences, pr. akademika Semenova 1-A, 142432
	Chernogolovka, Russia} \affiliation{National Research University
	Higher School of Economics, Myasnitskaya ul. 20, 101000 Moscow,
	Russia.}

\author{Elena~S.~Pikina}
\email[Corresponding author: ]{elenapikina@itp.ac.ru}
\affiliation{Landau Institute for
	Theoretical Physics, Russian Academy of Sciences, pr. akademika
	Semenova 1-A, 142432 Chernogolovka, Russia}
\affiliation{Institute of Solid State Physics, Russian Academy of
	Sciences, Academician Ossipyan str. 2, 142432 Chernogolovka, Russia}

\author{Efim~I.~Kats}
\affiliation{Landau Institute for Theoretical Physics, Russian
	Academy of Sciences, pr. akademika Semenova 1-A, 142432
	Chernogolovka, Russia}

\author{Noel~A.~Clark}
\affiliation{Department of Physics, University of Colorado Boulder, Boulder, CO 80309, USA}
\affiliation{Soft Materials Research Center, University of Colorado Boulder, Boulder, CO 80309, USA}

\author{Michael~Sprung}
\affiliation{Deutsches Elektronen-Synchrotron DESY, Notkestra{\ss}e 85, D-22607 Hamburg, Germany}

\author{Boris~I.~Ostrovskii}
\email[Corresponding author: ]{ostrenator@gmail.com}
\affiliation{Federal Scientific
	Research Center ``Crystallography and photonics'', Russian Academy
	of Sciences, Leninskii prospect 59, 119333 Moscow, Russia}
\affiliation{Institute of Solid State Physics, Russian Academy of
	Sciences, Academician Ossipyan str. 2, 142432 Chernogolovka, Russia}

\author{Ivan~A.~Vartanyants}
\email[Corresponding author: ]{ivan.vartaniants@desy.de}
\affiliation{Deutsches Elektronen-Synchrotron DESY, Notkestra{\ss}e 85, D-22607 Hamburg, Germany}
\affiliation{National Research Nuclear University MEPhI (Moscow Engineering Physics Institute), Kashirskoe shosse 31, 115409 Moscow, Russia}

    \date{\today}

    \begin{abstract}
		
	Experimental and theoretical studies of a smectic-hexatic transition in freely suspended films of 54COOBC compound are presented.
	X-ray investigations revealed a discontinuous first-order transition into the hexatic phase.
	Moreover, the temperature region of two phase coexistence near the phase transition point diminishes with film thickness.
	The coexistence width dependence on film thickness was derived on the basis of the Landau mean-field theory in the vicinity of the tricritical point (TCP).
	Close to TCP the surface hexatic ordering penetrates anomalously deep into the film interior.

\end{abstract}

\maketitle

Phase transitions are one of the richest and most intriguing phenomena in modern physics.
They are ubiquitous both in traditional condensed matter physics and other disciplines as diverse as biology
(phase transitions in lipid membranes \cite{Mirrink2005}),
astrophysics (transitions in dust plasma \cite{Petrov2015}) or cosmology (Kibble cosmological model \cite{Lee2015}).
Despite the great number of general work on phase transitions and their specific applications
(see, for example, classical books \cite{Chaikin, Fultz2014}),
the topic is still full of challenging open questions.
Recently, there has been an increased interest in fascinating material properties of the systems in the vicinity of the tricritical point (TCP), where the phase transition behavior changes from first to second order.
This was mainly followed by the discovery of fluctuation induced TCP in skyrmionic magnetic lattices \cite{Bauer2013, Garst2017,Mulkers2018} (although obvious connection exists with many other physical systems ranging from colloidal crystals \cite{Bolhuis1994} or block-copolymers \cite{Slyk2014,Lee2017,Mao2018} to helimagnets  \cite{Janoschek2013}).
Understanding of phase transition physics close to TCP is important in such systems as liquid water \cite{Zhang2011} or ice \cite{Himoto2014}.

Here we report on the behavior of the first-order smectic-A (Sm-A) -- hexatic-B (Hex-B) phase transition in liquid crystals (LCs) and find that it can be tuned to TCP by film thickness variation.
The Hex-B is a three-dimensional (3D) analogue of the common hexatic
phase \cite{Nelson2002,Kosterlitz2016,Agosta2018}.
It can be considered as a stack of parallel molecular layers, in which elongated molecules
are oriented on average along the layer normals, exhibiting long-range bond-orientational (BO) order and short-range positional order within each layer \cite{Pindak1981, Brock1986, Stoebe1995}.

Despite three decades of intensive studies, understanding of the Sm-A--Hex-B phase transition is still limited not only in details but even conceptually.
According to the Landau theory of phase transitions \cite{Landau1980,deGennes93}, this
transition is characterized by the two-component BO order parameter $\psi = |\psi |\exp (i 6\phi )$ (modulus and phase) and therefore the continuous phase transition must follow
the universal behavior predicted for such a case.
In reality, a number of experiments \cite{Stoebe1995, Jin1996, Haga1997, Jeu2003, Zaluzhnyy2017} do not support this concept demanding a revision of this simple picture. This has become especially important recently as experiments (x-rays and calorimetry) grow in
resolution and sophistication.

Interestingly, for 54COOBC
(\textit{n}-pentyl-4$'$-\textit{n}-pentanoyloxy-biphenyl-4-carboxylate)
LC compound the type of phase transition depends on dimensionality of the sample.
A first-order Sm-A--Hex-B phase transition occurs in bulk samples \cite{Jin1995, Jin1996}, while very thin freely suspended films (FSF) exhibit a continuous
Sm-A--Hex-B transition \cite{Chou1996,Chou1998}.
It was argued in Ref. \cite{Chou1998} that smectic--hexatics phase
transition in two-layered 54COOBC film occurrs via intermediate
Sm-A$'$ phase, which is characterized by the absence of BO order
and increased value of in-layer positional correlation length as
compared to common Sm-A phase.
In bulk samples the first-order Sm-A--Hex-B phase transition lies in the vicinity of the TCP.
We show that near the TCP the surface ordering penetrates anomalously deep into the interior of the film, which essentially influences the Sm-A--Hex-B phase transition even for thick films, consisting of thousands of layers.


X-ray studies were performed using 13 keV photons at the coherence beamline P10 of PETRA III synchrotron source at DESY.
The freely suspended films of 54COOBC compound were positioned perpendicular to
the incident x-ray beam, which was focused by compound refractive lenses to the size of about 2$\times$2 \SI{}{\micro\metre}$^2$ at full width at half maximum (FWHM).
At each temperature the film was scanned with x-ray beam over an area 100$\times$100 \SI{}{\micro\metre}$^2$ with a step of \SI{5}{\micro\metre} (for details of the experimental setup see \cite{Zaluzhnyy2017} and Appendix~\ref{AppA}).

\begin{figure}
	\includegraphics[width=0.8\linewidth]{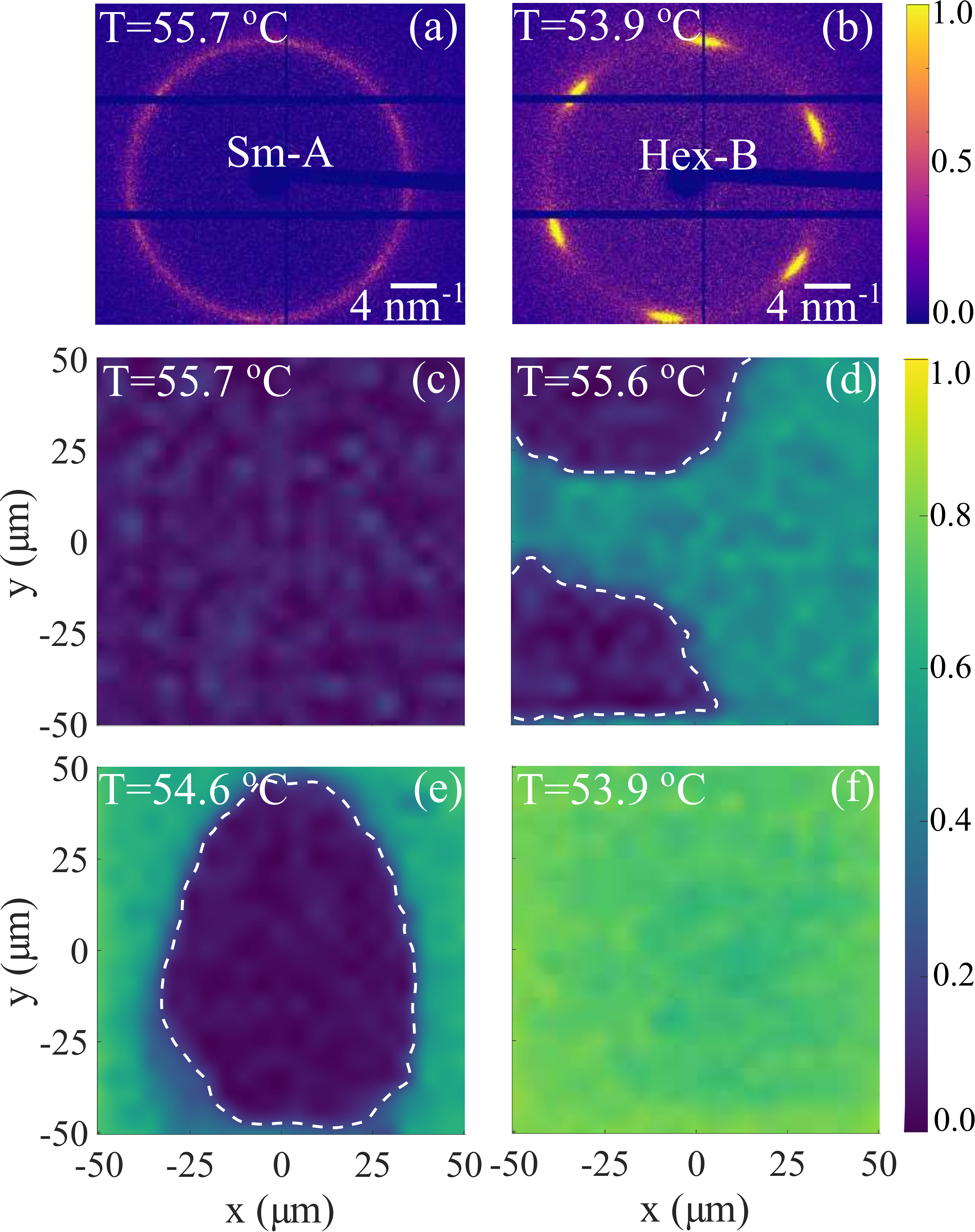}
	\caption{
		(a-b) Examples of diffraction patterns from 10 $\mu$m thick film in smectic
		(a) and hexatic (b) phases of 54COOBC.  Each image is averaged over 16 diffraction patterns collected
		within the area of 20$\times$20 \SI{}{\micro\metre}$^2$ for better visibility.
		Color refers to the normalized intensity of the scattered x-rays. (c-d) Spatially resolved maps
		of BO order parameter $C_6$ in 100$\times$100 \SI{}{\micro\metre}$^2$ region of 54COOBC film in the smectic phase (c), mixed state (d-e) and hexatic phase (f).
		Color indicates the local value of $C_6$, blue (dark) color corresponds to the Sm-A phase,
		and green (bright) corresponds to Hex-B. Dashed white line marks a border between the Sm-A and Hex-B phases in (d-e).
		\label{Figure1}
	}
\end{figure}

The FSFs of different  thickness ranging from \SI{2}{\micro\metre} to \SI{10}{\micro\metre} were measured on cooling and heating to observe formation of the hexatic phase at the Sm-A--Hex-B phase
transition.
Examples of the measured diffraction patterns in the  Sm-A and Hex-B phases are shown in Figs. \ref{Figure1}(a)-\ref{Figure1}(b).
The diffraction pattern in the Sm-A phase (Fig. \ref{Figure1}(a)) shows typical for liquids broad scattering ring centered at a scattering vector $q_0=4\pi/a\sqrt{3}\approx14$ nm$^{-1}$,
where $a\approx0.5$ nm is the average in-plane intermolecular distance.
In the Hex-B phase (Fig. \ref{Figure1}(b)) one can readily see sixfold modulation of the in-plane scattering, which is an evidence of the developing BO order.

To describe quantitatively magnitude of the angular  modulation of intensity in the Hex-B phase we used the BO order parameter $C_6$, which is defined as a normalized amplitude of the sixth-order angular Fourier component of the azimuthal scattered
intensity \cite{HigherComponents, Brock1986, Kurta2013}.
The value of $C_6$ can be conveniently determined using the angular x-ray
cross-correlation analysis (XCCA), which allows one to evaluate $C_6$ directly from the measured x-ray diffraction patterns \cite{Wochner2009, Kurta2013, Zaluzhnyy2015, Kurta2016} (see also Appendix~\ref{AppB}).
In the Sm-A phase the value of $C_6$ is equal to zero, while it approaches unity for ideal BO order \cite{Brock1986}.

Utilizing a microfocused x-ray beam, the spatially  resolved maps were retrieved
to reveal spatial variation of the BO order parameter $C_6$ within the scanned area.
These maps for \SI{10}{\micro\metre} thick FSF for different temperatures while cooling are shown in Figs. \ref{Figure1}(c)-\ref{Figure1}(f).
At high temperature (Fig. \ref{Figure1}(c)) the whole FSF is in the smectic phase, however, at lower temperatures the film gets non-uniform.
The coexistence of the Sm-A and Hex-B phases can be clearly seen  in Figs. \ref{Figure1}(d) and \ref{Figure1}(e).
The Hex-B phase co-exists with the Sm-A phase (Fig. \ref{Figure1}(d)) and then, at even lower temperatures, Hex-B becomes dominant and the Sm-A phase exists in the form of regions surrounded by the Hex-B phase (Fig. \ref{Figure1}(e)).
The size, shape and position of these regions may change when temperature varies, however we always observed the Sm-A phase surrounded by the Hex-B phase in different films of 54COOBC compound both on cooling and heating.
This observation can be explained by the fact that above the bulk Sm-A--Hex-B phase transition temperature, the hexatic order is first formed at the surface of the film and it penetrates into the inner layers on cooling \cite{Jin1996,Jeu2003,Stoebe1992}.
For such mechanism of the Hex-B phase formation, appearance of the Sm-A regions is favorable, contrary to the nucleation process, for which islands of Hex-B phase surrounded by the Sm-A phase should be observed.
During further cooling the whole FSF turns to the Hex-B phase with formation of single hexatic domains of a lateral size of hundreds of microns (Fig.
\ref{Figure1}(f)).

\begin{figure}
	\includegraphics[width=0.8\linewidth]{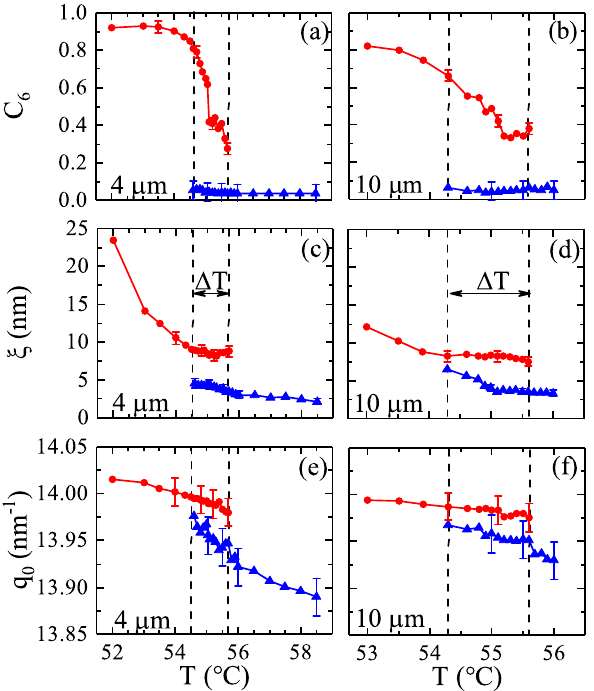}
	\caption{
		Temperature dependence of the BO order parameter $C_6$
		(a,b), the positional correlation length $\xi$ (c,d),
		and the scattering peak maximum position $q_0$ (e,f) in the Sm-A phase
		(blue triangles) and Hex-B phase (red circles) close to the region of  two phases coexistence.
		Data are shown for \SI{4}{\micro\metre} (left column) and \SI{10}{\micro\metre} (right column) thick films of 54COOBC.
		Dashed vertical lines indicate temperature region $\Delta T$ of two phases coexistence. Error bars are shown for each fifth experimental point.
		\label{Figure2}
	}
\end{figure}

\begin{figure}
	\includegraphics[width=0.8\linewidth]{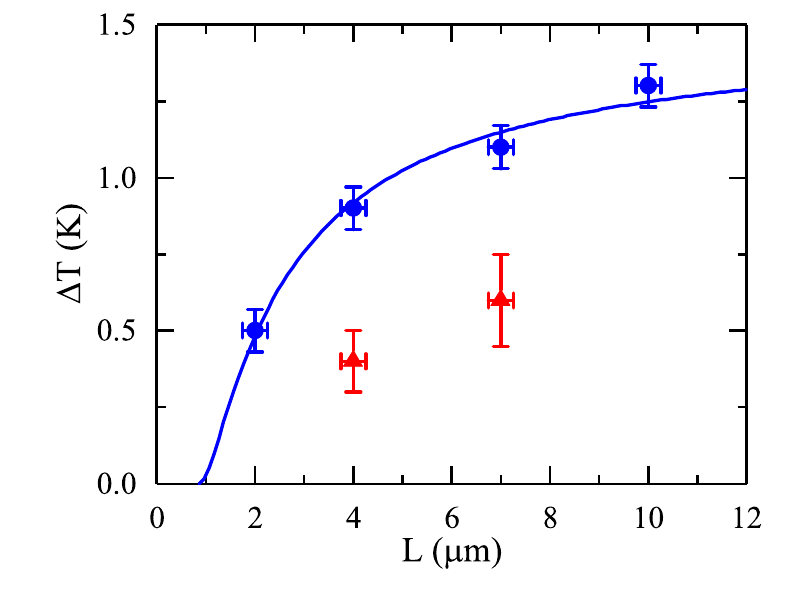}
	\caption{
		Temperature range $\Delta T$ of the Sm-A and Hex-B phase coexistence (two-phase region) as a function of the film
		thickness $L$. Blue circles correspond to data taken on cooling and red triangles – on heating of the
		54COOBC films.
		Fitting of the data on cooling by Eq. (\ref{surf8}) is shown by solid line.
		\label{Figure3}
	}
\end{figure}

We performed detailed analysis of scattering from the Hex-B and Sm-A regions of the film independently. In  Figs. \ref{Figure2}(a)-\ref{Figure2}(b) the temperature dependence of the BO order parameter $C_6$ for Sm-A (blue triangles) and Hex-B (red circles) phases is shown.
This dependence was obtained by averaging the local values of $C_6$ calculated for
each measured diffraction pattern over the regions of Sm-A and Hex-B phases separately (Figs. \ref{Figure1}(c)-\ref{Figure1}(f)).
In the Sm-A phase the magnitude of $C_6$ is vanishingly small and does not change with
the temperature.
In the Hex-B phase the value of $C_6$ rises during cooling, which corresponds to increase of the BO order in the low-temperature hexatic phase.
As it is expected for the first-order phase transition the BO order parameter does not change continuously from zero, but instead shows a discontinuous jump of the magnitude of  about 0.3 at the temperature when the first areas of the hexatic phase appear in the film.
Similar behavior was observed for the in-plane positional correlation length $\xi$ \cite{NoteSmA} (Figs. \ref{Figure2}(c)-\ref{Figure2}(d)) and position of the maximum of scattered intensity $q_0$, indicating an abrupt density change of about 0.4{\%} at the phase transition region (see Figs. \ref{Figure2}(c)-\ref{Figure2}(f) and Appendix~\ref{AppC}).
The observed discontinuities unambiguously indicate the first-order character of the
Sm-A--Hex-B phase transition in 54COOBC films.


An important outcome of our experiment was not only observation of the coexistence of the Sm-A and Hex-B phases in the finite temperature range $\Delta T$ but also revealing its dependence on the film thickness (blue circles in Fig. \ref{Figure3}).
We found that the width of the two-phase region is about 1.3 K for thick (\SI{10}{\micro\metre}) films and decreases for thinner films, reaching the value of about 500 mK for a film with a thickness of \SI{2}{\micro\metre}.
These observations are in good agreement with data reported for approximately \SI{0.25}{\micro\metre} thick film (100 molecular layers) of 54COOBC compound, in which coexistence of Sm-A and Hex-B phases was estimated to be within 90 mK \cite{Jin1995,Jin1996}.
Coexistence of two phases indicates that none of the phases (hexatic or smectic) is able to support the optimal  density of the film, and the compromise is achieved by two-phase equilibrium.
Another important observation arising from our experiment is that the
value of two-phase region $\Delta T$ obtained on cooling is larger
than on heating (red triangles in Fig. \ref{Figure3}). Such behavior
looks natural due to the presence of the hexatic surface ordering
in the smectic phase.  Thus the smectic phase can not be
overcooled, but contrary to that the hexatic phase can be relatively easy overheated.

In the following we derive an analytical expression which models the coexistence width as a function of film thickness.
For theoretical analysis of our experimental results let us first determine
the general thermodynamical conditions of the equilibrium coexistence of the Sm-A and Hex-B phases in bulk
\begin{equation}
f_\mathrm{H}[T,n_\mathrm{H}]=f_{\mathrm{Sm}}[T,n_\mathrm{Sm}] \ ,
\quad
\mu_\mathrm{H}[T,n_\mathrm{H}]=\mu_\mathrm{Sm}[T,n_\mathrm{Sm}] \
, \label{cond1}
\end{equation}
where $f[T,n]$ is the free energy  of the phase with density $n$
(the number of molecules per unit volume), $\mu[T,n]=\partial
f/\partial n$ is the chemical potential, and indexes $\mathrm{H}$ and
$\mathrm{Sm}$ correspond to the Hex-B and Sm-A phases,
respectively.
We designate as $\varphi[T,n]$ the free energy density of the Sm-A phase per unit volume, $f_\mathrm{Sm}[T,n_\mathrm{Sm}]=\varphi[T,n_\mathrm{Sm}]$.
In the vicinity of the TCP  the free energy
density
$f_\mathrm{H}$ of the Hex-B phase (per unit volume)
can be written according to conventional mean field theory \cite{Landau1980,Patashinskii1979}
as
$f_\mathrm{H}[T,n_\mathrm{H}]=\varphi[T,n_\mathrm{H}]+g[\psi]$, where
\begin{equation}
g[\psi]=a \vert \psi \vert^2 -\lambda \vert \psi \vert^4/6
+\zeta \vert \psi \vert^6/90
\label{cond2}
\end{equation}
and $a$, $\lambda$ and $\zeta$ are Landau coefficients which depend
on $T$ and $n$. The coefficients $a$ and $\lambda $ are assumed to
be small (they vanish at the TCP)  and $\lambda>0$ for a
first-order transition. We consider a relatively narrow
temperature interval where the Hex-B and Sm-A phases coexist.
Therefore  the coefficients $\lambda$ and $\zeta$ remain
approximately constant in this interval, whereas  coefficient $a$
has a standard form  $a=\alpha\tau$, where $\alpha>0$ is constant and
$\tau=(T-T_0)/T_0$ \cite{Landau1980,Patashinskii1979}.
Coefficient $a$ vanishes at a certain temperature $T_0$
which is in the  vicinity of TCP  close to the temperature of bulk
phase transition.

We will ignore dependence of $\lambda$ and $\zeta$ on $n_\mathrm{H}$.
The order parameter $\psi$ in the hexatic phase is determined by minimization of expression (\ref{cond2}): $\vert
\psi_{\mathrm{H}}\vert^2=(5\lambda/\zeta)(1
+\sqrt{1-6a\zeta(5\lambda^2)^{-1}})$, which is  real  for $a<a_+$,
where $a_{+}=5\lambda^2/6\zeta$. This gives a condition of
existence of the hexatic phase. In turn, assuming a small
difference $\Delta n$ between  $n_\mathrm{Sm}$ and $n_\mathrm{H}$,
$n_\mathrm{H}=n_\mathrm{Sm}-\Delta n$, and minimizing
expression (\ref{cond2}) with respect to the order parameter
$\psi$, from Eqs. (\ref{cond1}) we obtain
\begin{align}
\Delta n&=\frac{\partial a}{\partial n}\, \frac{\partial
	n}{\partial \mu}  \,\vert\psi \vert^2 \ ,
\label{cond6} \\
\mu \frac{\partial a}{\partial n}\, \frac{\partial n}{\partial
	\mu} &=\frac{2a}{3}-\frac{\lambda}{18}\vert \psi \vert^2 \ ,
\label{cond7}
\end{align}
where $\mu=\partial \varphi/\partial n$.
Expression (\ref{cond7}) determines the compressibility of
the Sm-A phase at the equilibrium curve. Upon diminishing $a$ the
right-hand side of Eq. (\ref{cond7}) monotonically decreases and
turns to zero at $a_{-}=5\lambda^2 /  8\zeta$ (i.e.  at
$\tau_{c}=5\lambda ^{2} / (8\alpha\zeta)$, $\vert \psi_{c}
\vert^{2}=15\lambda / 2\zeta$). At this point $(\partial
n/\partial\mu)$  becomes zero, that implies instability of the
smectic phase. Thus, from condition $a_-<a<a_+$ one can find that
the equilibrium between Hex-B and Sm-A phases occurs in bulk in
the  interval
\begin{equation}
\label{cond9}
\Delta T =\frac{5\lambda^2}{24\alpha\zeta}\,T_0 \ .
\end{equation}
To take surface effects into account, one should include the
gradient term  into the Landau functional for the field $\psi$
(per unit area).
Then we obtain for the free energy of the film
\begin{equation}
\label{surf1} {\cal F}\,=\,S\,\int_{-L/2}^{ L/2} dz\, \big\{ b\,
(\partial_z {\psi})^2 \,+\,g[\psi] \big\} \ ,
\end{equation}
where   $b>0$, $z=\pm\,{ L/2}$ correspond to free surfaces of FSF
of thickness $L$ and surface area $S$.
We assume in the following
that the phase of the hexatic order parameter $\psi$ is fixed, and
therefore one can use real values of $\psi$. Due to  the symmetry
properties of FSF  we have condition $\partial_z\psi[0]=0$ in the
middle of the film. Minimization of Eq. (\ref{surf1}) with respect
to $\psi$ gives Euler-Lagrange equation, which can be integrated
once to yield
\begin{equation}
\label{profb} b (\partial_z \psi)^{2}= g[\psi]+ C_1 \ ,
\end{equation}
where $C_1=-g[\psi_m]$,
$\psi[0]\equiv\psi_m$. At $z\to 0$ the solution for $\psi[z]$ approaches its
asymptotic bulk value, i.e.  $\psi_m\approx\psi_\mathrm{H}$  in
hexatic phase  and $\psi_m\approx0$ in smectic phase.  In above we
have used the assumption $L\gg\xi_z$, which is true for the thick
FSF under consideration. Here  $\xi_z$ is a  correlation length
along $z$ axis,  which is much larger then molecular size close to
the TCP.

There are two contributions to the free energy of the film ${\cal F}$ of the film:
the bulk energy ${\cal F}^{(b)}=g[\psi_\mathrm{H}]\,LS$ and  surface
energy ${\cal F}^{(s)}$.
The last one can be found after subtracting
${\cal F}^{(b)}$ from expression (\ref{surf1}).
With the use of Eq. (\ref{profb}) we can obtain
\begin{eqnarray}
\label{surfj}
{\cal F}^{(s)}\,=\,2\,b\,S\,\int_{-L/2}^{ L/2}
dz\ (\partial_z \psi)^2 \ .
\end{eqnarray}
Assuming that $\psi^{(s)}\gg \psi_\mathrm{H}$, where $\psi^{(s)}$ is the surface value of $\psi$, it follows from Eqs. (\ref{cond2}) and (\ref{profb}) that in
both phases there exists a region near the surface where $\psi^{6}$  becomes a leading term on the right-hand side of Eq. (\ref{profb})  and we find: $ \psi^2\approx(\psi^{(s)})^2(1
+(\psi^{(s)})^2\,\sqrt{2\,\zeta/(45\,b)\,}\,(L/2-z))^{-\,1}$ (for
$z>0$).
The surface value of $\psi$ is identified for  Hex-B and Sm-A
phases as  $\psi^{(s)}_{\mathrm{H}}$ and $\psi^{(s)}_\mathrm{Sm}$,
respectively
($\psi^{(s)}_{\mathrm{H}}>\psi^{(s)}_\mathrm{Sm}$).

The analysis of solution of  Eq. (\ref{profb}) indicates
that even in the Sm-A phase  there is an  anomalous  penetration
of hexatic ordering in FSF close to TCP.
Substituting the above solution into Eq. (\ref{surfj}) one obtains main
contributions to the surface energy. Using expressions
(\ref{surfj}) and (\ref{profb}) one finds the difference between
the surface energies of the Hex-B and Sm-A phases
\begin{equation}
\label{DFs2}
\frac{\Delta{\cal F}^{(s)}}{4\sqrt{b}
	S}=\int_{\psi_\mathrm{H}}^{\psi^{(s)}_{\mathrm{H}}}d\psi
\sqrt{g[\psi]-g[\psi_\mathrm{H}]}-\int_0^{\psi^{(s)}_\mathrm{Sm}}d\psi
\sqrt{g[\psi]}
\, .
\end{equation}
Using  equation $\partial
g[\psi_\mathrm{H}]/\partial\psi_\mathrm{H}=0$, we obtain from Eq.
(\ref{DFs2})
\begin{equation}
\Delta{\cal F}^{(s)}=
(b\zeta)^{1/2}  w[\psi_\mathrm{H}^2 \zeta/\lambda]\,\psi_\mathrm{H}^4\, S \ ,
\label{surf77}
\end{equation}
where $w[\psi_\mathrm{H}^2 \zeta/\lambda]$ is a positive dimensionless
function which depends on the surface values of $\psi$ and can be
found  numerically  \cite{NextStep}.  Comparing  term
(\ref{surf77}) with ${\cal F}^{(s)}$ we conclude that  surface effects
produce
an effective positive correction   $\delta\lambda\simeq
6\,(b\zeta)^{1/2}\,w[\psi_\mathrm{H}^2 \zeta/\lambda]\,L^{-1}$ to
the coefficient $\lambda$. Using  Eq. (\ref{cond9}) we finally
arrive to
\begin{equation}
\label{surf8}
\Delta T
=\frac{5\lambda^2}{24\alpha\zeta}\,T_0\,\Big(1-\frac{L_0}{L}\Big)^2
\ ,
\end{equation}
where $L_0=L(\delta\lambda/\lambda)$ is a characteristic length
scale. Fitting of the experimental data with Eq. (\ref{surf8})
shows a good agreement with theoretical predictions (see Fig.
\ref{Figure3}), and gives  $L_0=0.9$~\SI{}{\micro\metre}.


In conclusion, we report on detailed spatially resolved x-ray studies of a first-order Sm-A--Hex-B phase transition in free standing films of 54COOBC of various thickness.
Microfocused X-ray diffraction in combination with XCCA technique allowed us
directly observe coexistence of Sm-A and Hex-B phases.
Experimentally measured temperature dependences of such structural parameters as $C_6$, $\xi$, and $q_0$ exhibit discontinuous behavior at transition temperature, which were not observed for second-order Sm-A--Hex-B phase transition in other compounds (see Appendix~\ref{AppD}).
We also found that the width of the two-phase region $\Delta T$ at the
Sm-A--Hex-B transition becomes narrower for thinner films,
reaching the value of about 500 mK for a \SI{2}{\micro\metre} thick film.
This indicates that the phase behavior of the 54COOBC films is strongly affected by the surface hexatic ordering field,
which penetrates to interior layers of the film over large distances induced by the proximity of the Sm-A--Hex-B transition in 54COOBC to a TCP.
Analytical expression for $\Delta T$ obtained from the Landau mean field theory is in a good agreement with the experimental data.
This gives a unique possibility to approach TCP at the Sm-A--Hex-B phase transition line by varying the film thickness and experimentally investigate general properties of the
phase transitions in the vicinity of TCP.
This new approach is quite general and can be applied to a large class of systems
exhibiting TCP, for example, helimagnetic films \cite{Janoschek2013}, or
recently discovered materials with skyrmionic magnetic lattices
\cite{Garst2017,Mulkers2018}.

\begin{acknowledgments}

	We acknowledge E. Weckert for fruitful discussions and support of the project, F. Westermeister for the support during the experiment, A.~R.~Muratov for theoretical discussions, and R. Gehrke for careful reading of the manuscript. The work of  R.~M.~Kh.,  E.~S.~P. and B.~I.~O. was  supported by the Russian Science Foundation (Grant No. 18-12-00108). N. A. C. acknowledges support of U.S. National Science Foundation Grant DMR1420736.
\end{acknowledgments}


\appendix

\section{Details of X-ray experiment}
\label{AppA}

\begin{figure}
	\includegraphics[width=0.8\linewidth]{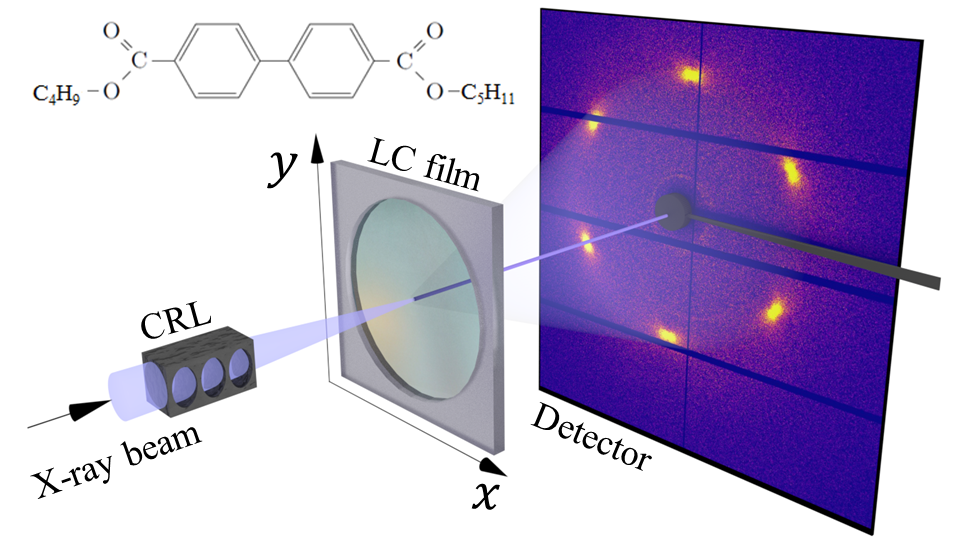}
	\caption{
		Schematic representation of x-ray diffraction setup used in the experiment. The x-ray beam is focused by compound refractive lenses (CRL). The free standing LC film is oriented perpendicular to the incoming beam, and piezoelectric motors allow to scan the film in XY-plane. The diffraction pattern is recorder in transmission geometry by Eiger 4M detector placed behind the LC film. The direct beam is blocked by a beamstop. Chemical structure of 54COOBC compound is shown in the inset.
		\label{Suppl_Setup}
	}
\end{figure}

In the present work we studied free standing films of liquid crystal (LC) compound \textit{n}-pentyl-4$'$-\textit{n}-pentanoyloxy-biphenyl-4-carboxylate (54COOBC) with chemical sturture shown in еру inset in Fig. \ref{Suppl_Setup}, which has the following phase sequence in bulk: I~(70~$\degree$C) Sm-A~(55~$\degree$C)  Hex-B~(53~$\degree$C) Cr-B \cite{Jin1995, Chou1996, Chou1998, Surendranath1985}.
The films were drawn across a circular hole of 2 mm in diameter in a thin glass plate \cite{Kurta2013,Zaluzhnyy2015,Zaluzhnyy2017}.
By varying the temperature and the speed of drawing, one can produce films of different thickness ranging from \SI{1}{\micro\metre} to \SI{10}{\micro\metre}.
The thickness of the films was measured using AVANTES fiber optical spectrometer.

The films were placed inside FS1 sample stage from INSTEC connected to mK1000 temperature controller. The accuracy of temperature control during experiment was about 0.005~$\degree$C.
X-ray studies were performed at the coherence beamline P10 of PETRA III synchrotron source at DESY.
The scheme of the experiment is shown in Fig. \ref{Suppl_Setup}.
The film was placed perpendicular to the incident X-ray beam, which was focused by compound refractive lenses to the size of about 2$\times$2 \SI{}{\micro\metre}$^2$ at full width at half maximum (FWHM) \cite{Zozulya2012}.
The energy of incident X-ray photons was 13 keV, which corresponds to wavelength of $\lambda=0.954$ \AA.
The scattering signal was recorded by EIGER 4M detector (2070$\times$2167 pixels of 75$\times$75 \SI{}{\micro\metre}$^2$ size) placed 232 mm behind the sample.
At each temperature the film was scanned with a micro-focused beam over an area 100$\times$100 \SI{}{\micro\metre}$^2$ with a step of \SI{5}{\micro\metre}.
Prior further analysis the collected diffraction patterns were corrected for background scattering and horizontal polarization of synchrotron radiation.
Each diffraction pattern was collected at exposure time of 0.5 s to avoid radiation damage of the film.
Also large scans (1$\times$1 mm$^2$ with \SI{25}{\micro\metre} step) were performed to investigate the spatial variation of the bond-ordientational order in the Hex-B phase at large scale.

\section{X-ray cross-correlation analysis}
\label{AppB}

\begin{figure}
	\includegraphics[width=0.45\linewidth]{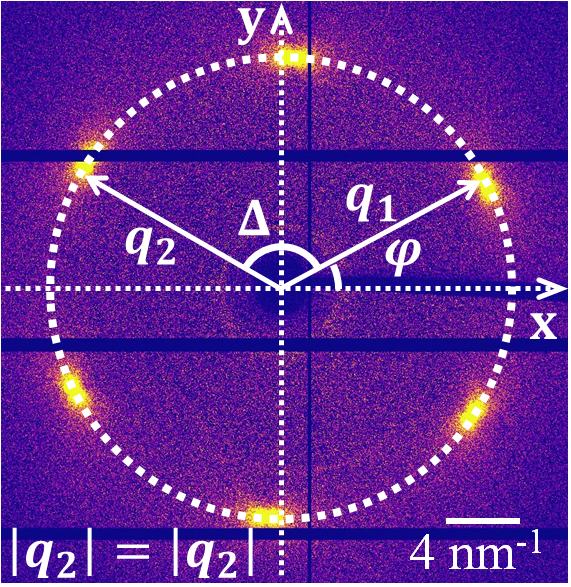}
\caption{
	Example of the diffraction patten from Hex-B phase and polar system of coordinates for evaluation of the angular cross-correlation function $G(q,\Delta)$.
	\label{Suppl_Datector}
}
\end{figure}

An angular X-ray Cross-Correlation Analysis (XCCA) is a technique that allows one to study local order and especially rotational symmetry present in a system by analysis of angular distribution of scattered intensity of the measured diffraction patterns \cite{Wochner2009}.
In this work XCCA is used for direct evaluation of the bond-orientational (BO) order parameters in the Hex-B phase \cite{Kurta2013}.
The key element of XCCA is two-point angular correlation function evaluated for each diffraction pattern \cite{Kurta2013a, Kurta2016}
\begin{equation}
\label{XCCA1}
G(q,\Delta)=\langle I(q,\varphi)I(q, \varphi+\Delta)\rangle_\varphi \, .
\end{equation}
Here $(q,\varphi)$ are polar coordinates at the detector plane, $\Delta$ is the angular variable, $\langle...\rangle_\varphi$ denotes averaging over azimuthal angle $\varphi$ (Fig. \ref{Suppl_Datector}).
Information about rotational symmetry of diffraction pattern contained in cross-correlation function (\ref{XCCA1}) can be easily approached by utilizing angular Fourier components
\begin{equation}
\label{XCCA2}
G_n(q)=\frac{1}{2\pi}\int_{0}^{2\pi}G(q,\Delta)\exp^{-in\Delta}d\Delta \, .
\end{equation}
It can be shown that the values of Fourier components $G_n(q)$ are directly related to the angular Fourier components of intensity: $G_n(q)=|I_n(q)|^2$ \cite{Kurta2013a}.
It is important to note, that angular cross-correlation function $G(q,\Delta)$ as well as its Fourier components $G_n(q)$ can be averaged over ensemble of diffraction patterns to obtain representative information about the sample and improve signal-to-noise ratio.
In contrast, one cannot average individual diffraction patterns or Fourier components of intensity $I_n(q)$, because the resulting averaged diffraction pattern can be isotropic, that will lead to loss of information about the orientational order contained in individual diffraction patterns.

For Hex-B phase the bond-orientational (BO) order parameters $C_{6m}$ ($m$ is an integer), are defined as normalized magnitude of sixfold angular Fourier components of intensity $I_{6m}(q_0)$ at the maximum of the scattered intensity momentum transfer $q_0$.
They can be directly evaluated through the Fourier components of cross-correlation function $G_{6m}(q_0)$ \cite{Kurta2013, Zaluzhnyy2015, Zaluzhnyy2017}
\begin{equation}
\label{XCCA3}
C_{6m}=\Bigg|\frac{I_{6m}(q_0)}{I_0(q_0)}\Bigg|=\sqrt{\frac{G_{6m}(q_0)}{G_0(q_0)}}.
\end{equation}
By definition the values of the BO order parameters are normalized, $0\le C_{6m} \le 1$; in the Sm-A phase $C_{6m}=0$ for all integer $m$, while in the Hex-B phase the BO order parameters $C_{6m}$ successively  attains certain values upon temperature decrease \cite{Brock1989,Zaluzhnyy2015,Zaluzhnyy2017}.
In this work we analyzed temperature dependence of the fundamental BO order parameter $C_6$, which is sufficient to distinguish Sm-A and Hex-B phases, while analysis of higher components $C_{6m}$ providing more detailed information of BO order will be a subject of further publication.

Another advantage of XCCA is the ability of this method to handle experimental noise, which is always present in diffraction patterns. In ideal case, in Sm-A phase only zero-order Fourier component $G_0(q_0)$ has positive value, while all other components are equal to zero $G_{n\ne0}(q_0)=0$ due to absence of the BO order.
However, for real experimental data all Fourier components $G_n(q_0)$ for $n>0$ have small non-zero values even in Sm-A phase due to noise.
Thus, theoretical criterion $C_6=0$ in Sm-A and $C_6>0$ in Hex-B does not work for real experimental data.

In order to overcome this problem, a statistical analysis of Fourier components $C_n(q_0)$ was performed, using diffraction patterns measured at different positions of LC film.
It turned out, that the the fundamental BO order parameter $C_6$ in the uniform Sm-A phase (measured at high temperature $T=59$ $\degree$C) has mean value of $\langle C_6 \rangle=0.05$ and standard deviation $\Delta C_6=0.04$.
Based on this analysis, a threshold value $C_t=\langle C_6\rangle+\Delta C_6=0.09$ was introduced to separate Sm-A and Hex-B phases.
Thus, each measured diffraction pattern at any temperature was attributed to one or another phase by following criterion: $C_6\ge0.09$ for Hex-B and $C_6<0.09$ for Sm-A.
This criterion was used throughout the work to perform separate analysis of Sm-A and Hex-B phases.

\begin{figure}
	\includegraphics[width=0.8\linewidth]{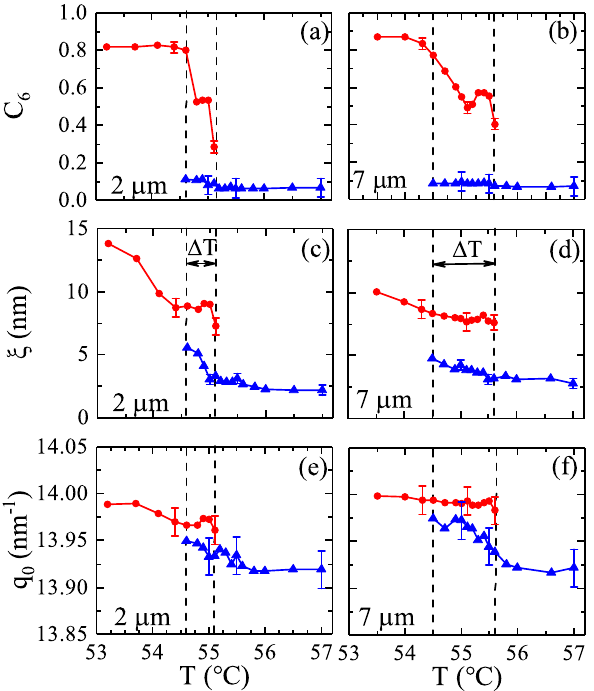}
	\caption{
		Temperature dependence of the BO order parameter $C_6$
		(a,b),  the positional correlation length $\xi$  (c,d),
		and the scattering peak maximum position $q_0$ (e,f) in the Sm-A phase
		(blue triangles) and Hex-B phase (red circles) close to the region of  two phases coexistence.
		Data are shown for \SI{2}{\micro\metre} (left column) and \SI{7}{\micro\metre} (right column) thick films of 54COOBC (compare with results for  \SI{4}{\micro\metre} and \SI{10}{\micro\metre} thick films in Fig. 2 of the main text).
		Dashed vertical lines indicate temperature region $\Delta T$ of two phases coexistence. Error bars are shown for each fifth experimental point.
		\label{Suppl_OtherThickness}	
	}
\end{figure}

Temperature dependence of the BO order parameter $C_6$ is shown for  \SI{4}{\micro\metre} (left) and \SI{10}{\micro\metre} thick films in Figs. 2(a)-2(b) of the main text and for \SI{2}{\micro\metre} (left) and \SI{7}{\micro\metre} thick films in Figs. \ref{Suppl_OtherThickness}(a)-\ref{Suppl_OtherThickness}(b).
In these figures one can clearly see the difference between value of $C_6$ in the Sm-A and Hex-B phases for 54COOBC compound.

\section{Positional correlation length and in-layer density}
\label{AppC}

One of the important characteristics of the in-plane short-range order in smectic  and hexatic phases is the positional correlation length $\xi$, determining the length scale over which the positional correlations between the molecules decay \cite{Zaluzhnyy2016}.
In the Hex-B phase the value of positional correlation length can be calculated as $\xi=1/\Delta q$, where $\Delta q$ is a half width at half maximum (HWHM) of the radial cross-section of the  hexatic diffraction peak through its maximum.
In this work we evaluated $\Delta q$ by fitting the radial intensity profile with Lorentzian function \cite{Zaluzhnyy2015,Zaluzhnyy2016}.

The temperature dependence of the positional correlation length averaged over regions of Sm-A and Hex-B phases is shown in Figs. 2(c)-2(d) of the main text for \SI{4}{\micro\metre} and \SI{10}{\micro\metre} thick LC films and in Figs. \ref{Suppl_OtherThickness}(c)-\ref{Suppl_OtherThickness}(d) for \SI{2}{\micro\metre} and \SI{7}{\micro\metre} thick films.
In the Sm-A phase the value of $\xi$ gradually increases from approximately 2.5 nm at $T=57$~$\degree$C to about 5 nm at the lowest temperature of the Sm-A phase existence.
The discontinuity in the $\xi$ values at the borders of two phase region is a prime indication of the first-order character of the Sm-A--Hex-B transition in 54COOBC films. These abrupt changes of correlation length unambiguously determine the range of coexistence of the Sm-A and Hex-B phases in both films.
In the Hex-B phase the positional correlation length further increases on cooling up to the value of 23 nm until the crystal phase is formed.
Such a behavior is attributed to coupling between the BO order and positional correlations in the Hex-B phase \cite{Zaluzhnyy2015,Aeppli1984}.

Qualitatively the growth of $\xi$ within the two-phase region observed in our experiment  corresponds to the range of enhanced values of the positional correlations of Sm-A$'$ phase reported earlier for a two-layer LC films of 54COOBC \cite{Chou1996,Chou1998}.
Use of the large focus size of the electron beam (\SI{50}{\micro\metre} in diameter) in this work did not allow spatial separation of Sm-A and Hex-B phases, which may lead to averaging over two different phases.
Indeed, using x-ray beam focused down to \SI{2}{\micro\metre} (FWHM) we showed that characteristic length scale of Sm-A phase regions within the temperature interval $\Delta T$ of two phases coexistence is of the order of \SI{50}{\micro\metre}.
Thus, one truly needs focused beam for direct observation of two phases in freely suspended films.

Another important parameter characterizing the Sm-A--Hex-B phase transition is the position of the maximum of scattered intensity $q_0$ in the smectic and hexatic phases. The value of $q_0$ is inversely proportional to the average in-plane separation between LC molecules, and thus indicates the variation of density across the transition point. According to the theory of Aeppli and Bruinsma \cite{Aeppli1984} the growing of fluctuations of the BO order parameter in the vicinity of a second order Sm-A--Hex-B phase transition leads to a continuous increase of the peak’s maximum position $q_0$ and the appearance of inflection point of $q_0(T)$ at phase transition temperature.
This is indeed observed for many LC compounds possessing a second-order Sm-A--Hex-B phase transition \cite{Zaluzhnyy2017, Zaluzhnyy2015, Davey1984}.

Situation becomes different for the first-order Sm-A--Hex-B phase transition, where one can expect a discontinuous density jump at the phase transition point.
In Figs. 2(e)-2(f) of the main text the temperature dependence of $q_0$ averaged separately over the regions of Sm-A and Hex-B phases is shown for \SI{4}{\micro\metre} and \SI{10}{\micro\metre} thick films of 54COOBC compound and in Figs. \ref{Suppl_OtherThickness}(e)-\ref{Suppl_OtherThickness}(f) for \SI{2}{\micro\metre} and \SI{7}{\micro\metre} thick films.
The discontinuity in $q_0$ at the borders of two phase region is readily seen, thus providing one more firm evidence for the first-order character of the Sm-A to Hex-B transition in 54COOBC films.
The values of $q_0$ in both coexisting phases differ by about 0.4\%, which is rather small for the structural phase transitions of the first order and can be explained by closeness of the system to the tricritical point (TCP).

\section{Comparison with other compounds}
\label{AppD}

	\begin{figure}
	\includegraphics[width=0.45\linewidth]{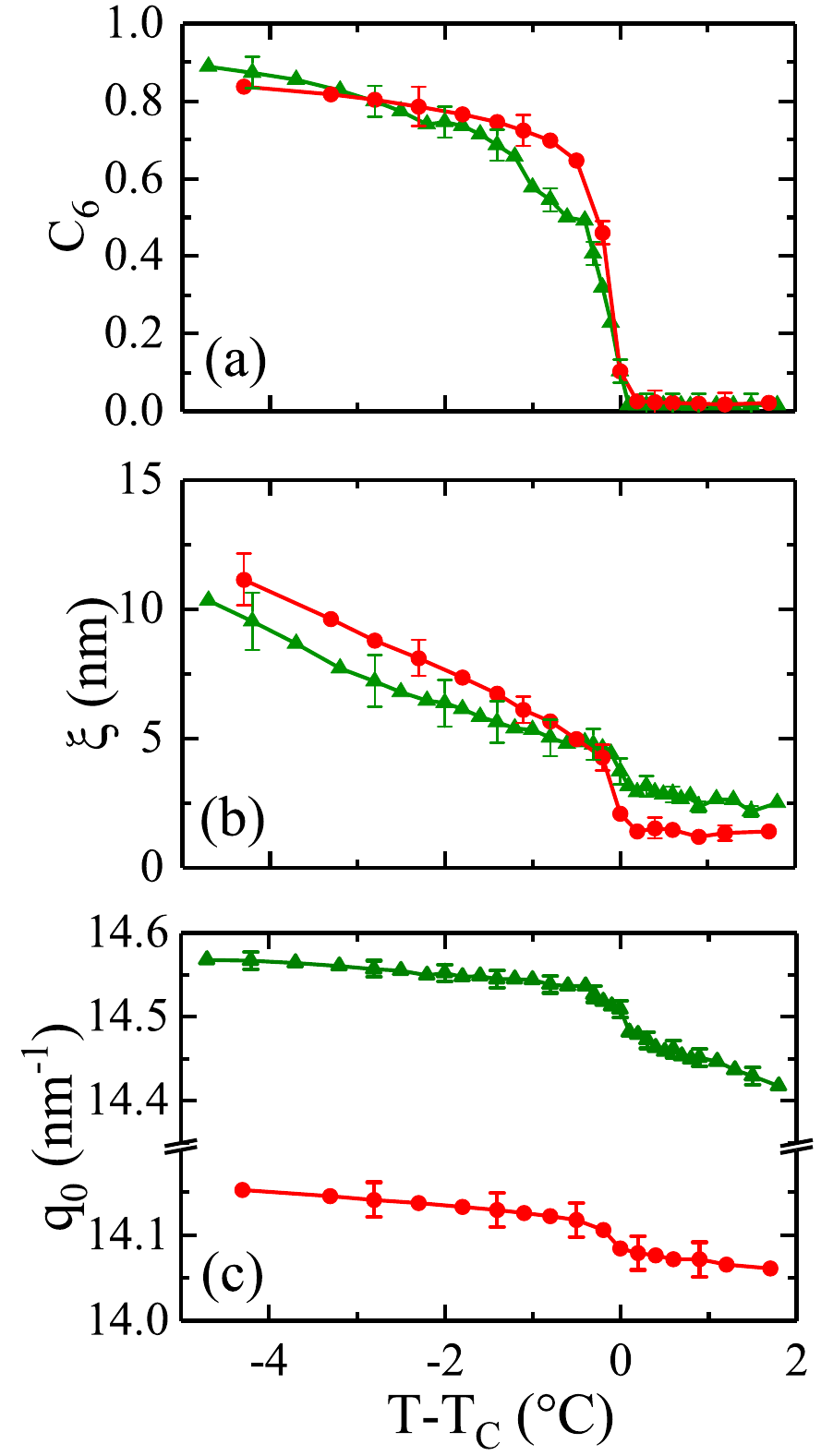}
	\caption{
		Temperature dependence of positional correlation length $\xi$ (a), position of the maximum of scattered intensity $q_0$ (b), and fundamental BO order parameter $C_6$ (c) for \SI{9}{\micro\metre} thick PIRO6 film (green triangles) and \SI{4}{\micro\metre} thick 3(10)OBC film (red circles).
		For convenience we used relative temperature $T-T_C$ measured with respect to the second-order Sm-A -– Hex-B phase transition temperature $T_C$, which is 66.3 $\degree$C for 3(10)OBC and 96.2 $\degree$C for PIRO6 \cite{Zaluzhnyy2017}.
		\label{Suppl_PIRO}	
	}
\end{figure}

Contrary to 54COOBC, the phase behavior of 3(10)OBC \cite{Stoebe1992,Stoebe1992a, Zaluzhnyy2015, Zaluzhnyy2017} and PIRO6 \cite{Pyzuk1995, Zaluzhnyy2017} LCs studied by us earlier is of the conventional second order with a continuous change of parameters  C$_6$, $\xi$, and $q_0$ through the Sm-A--Hex-B transition.
In Fig. \ref{Suppl_PIRO} the results of similar x-ray diffraction experiments conducted under the same conditions and using the same equipment on \SI{4}{\micro\metre} thick 3(10)OBC and \SI{9}{\micro\metre} thick PIRO6 films are displayed.
The experiment with PIRO6 compound was performed just after measurements of 54COOBC compound presented above, and 3(10)OBC compounds was studied in 2014 using the same setup and similar conditions (see \cite{Zaluzhnyy2015, Zaluzhnyy2016, Zaluzhnyy2017, Zaluzhnyy2017a} for results of these investigations).
There is no coexistence of two phases for these compounds, which was the case for the first-order Sm-A--Hex-B phase transition in 54COOBC.
The BO order parameter $C_6$, positional correlation length $\xi$, and maximum of scattered intensity $q_0$ vary continuously across the phase transition region.
These findings not only prove the accuracy of experiment but also indicate that thermodynamic behavior of LC films may be significantly different in the first- and second-order phase transition case, even if in both cases they occur in the vicinity of the TCP.

\newpage
\bibliography{References_54COOBC_arxive}

\begin{thebibliography}{49}
\bibitem [{Nex()}]{NextStep}%
  \BibitemOpen
  \href@noop {} {}\bibinfo {howpublished} {The full theory will be the subject
  of a separate publication.}\BibitemShut {Stop}%
\end{thebibliography}

\end{document}